\documentclass[sigconf]{acmart}

\usepackage{graphicx}
\usepackage{subfigure}
\AtBeginDocument{%
  \providecommand\BibTeX{{%
    \normalfont B\kern-0.5em{\scshape i\kern-0.25em b}\kern-0.8em\TeX}}}

\setcopyright{acmcopyright}
\copyrightyear{2019}
\copyrightyear{2019} 
\acmYear{2019} 
\acmConference[SIGSPATIAL '19]{27th ACM SIGSPATIAL International Conference on Advances in Geographic Information Systems}{November 5--8, 2019}{Chicago, IL, USA}
\acmBooktitle{27th ACM SIGSPATIAL International Conference on Advances in Geographic Information Systems (SIGSPATIAL '19), November 5--8, 2019, Chicago, IL, USA}
\acmDOI{10.1145/3347146.3359079}
\acmISBN{978-1-4503-6909-1/19/11}

\begin{document}

\title{Flight Delay Prediction using Airport Situational Awareness Map}

\author{Wei Shao}
\authornote{Both authors contributed equally to this research.}
\email{wei.shao@rmit.edu.au}
\affiliation{%
  \institution{RMIT University}
  \city{Melbourne}
  \state{VIC}
  \country{Australia}
  \postcode{3000}
}

\author{Arian Prabowo}
\authornotemark[1]
\email{arian.prabowo@rmit.edu.au}
\affiliation{%
  \institution{RMIT University}
  \institution{Data61/CSIRO}
  \city{Melbourne}
  \state{VIC}
  \country{Australia}
}

\author{Sichen Zhao}
\email{sichen.zhao@rmit.edu.au}
\affiliation{%
  \institution{RMIT University}
  \city{Melbourne}
  \state{VIC}
  \country{Australia}
  \postcode{3000}
}

\author{Siyu Tan}
\email{siyutan.cici@gmail.com}
\affiliation{%
  \institution{Xi'an University of Technology}
  \city{Xi'an}
  \state{Shaanxi}
  \country{China}}

\author{Piotr Koniusz}
\email{piotr.koniusz@data61.csiro.au}
\affiliation{%
  \institution{Data61/CSIRO}
  \institution{Australian National University}
  \city{Canberra}
  \state{ACT}
  \country{Australia}
}

\author{Jeffrey Chan}
\email{jeffrey.chan@rmit.edu.au}
\affiliation{%
  \institution{RMIT University}
  \city{Melbourne}
  \state{VIC}
  \country{Australia}
  \postcode{3000}
}

\author{Xinhong Hei}
\email{heixinhong@xaut.edu.cn}
\affiliation{%
  \institution{Xi'an University of Technology}
  \city{Xi'an}
  \state{Shaanxi}
  \country{China}}

\author{Bradley Feest}
\email{Bradley.Feest@ngc.com}
\affiliation{%
  \institution{Northrop Grumman Corporation}
  \city{Redondo Beach}
  \state{CA}
  \country{USA}}
  
\author{Flora D. Salim}
\email{flora.salim@rmit.edu.au}
\affiliation{%
  \institution{RMIT University}
  \city{Melbourne}
  \state{VIC}
  \country{Australia}
  \postcode{3000}
}
\renewcommand{\shortauthors}{Shao, et al.}

\begin{abstract}
The prediction of flight delays plays a significantly important role for airlines and travellers because flight delays cause not only tremendous economic loss but also potential security risks. In this work, we aim to integrate multiple data sources to predict the departure delay of a scheduled flight. Different from previous work, we are the first group, to our best knowledge, to take advantage of airport situational awareness map, which is defined as airport traffic complexity (ATC), and combine the proposed ATC factors with weather conditions and flight information. Features engineering methods and most state-of-the-art machine learning algorithms are applied to a large real-world data sources. We reveal a couple of factors at the airport which has a significant impact on flight departure delay time. The prediction results show that the proposed factors are the main reasons behind the flight delays. Using our proposed framework, an improvement in accuracy for flight departure delay prediction is obtained.
\end{abstract}





\maketitle

\section{Introduction}
In the age of globalisation, air travel has become a major method for long distance international and domestic travelling everyday around the world. With the increasing number of flights and airlines, flight delay has become a severe problem for airline companies and air travel passengers. In 2017, the Federal Aviation Administration (FAA) estimated the annual cost of flight delays to be \$26.6 billion in the United States. In 2018, the United States Government Accountability Office (GAO) reported that flight delay and cancellation accounted for an average of almost 33 percent of complaints from air travel passengers for selected airlines \cite{GAO19}. Flight delays not only cause tremendous economic costs but also bad psychological impact on air travellers.

Flight delays are widely spread in air travel area. In recent years, around a quarter of all commercial flights have been delayed or cancelled \cite{GAO11}. Among a variety of flight delay types, tarmac delay is the most common and intolerable because passengers need to wait for those delayed flights at the airport or in the aeroplane without information about how long they need to wait. Around 1600 flights have suffered a tarmac delay greater than three hours during the summertime between 2004 to 2010 \cite{GAO11}.

\begin{figure}
\includegraphics[width=\linewidth]{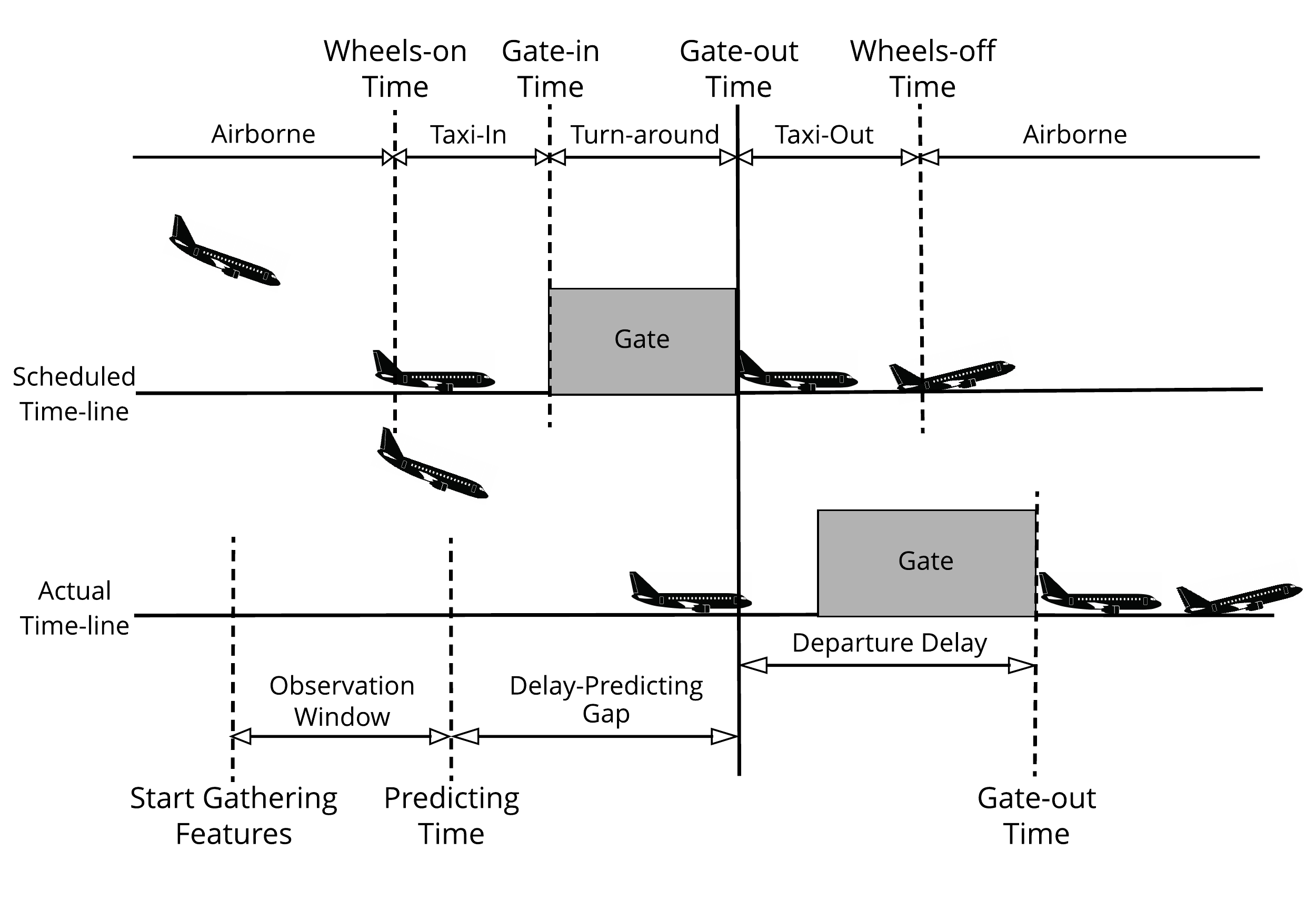}
\caption{The demonstration of our task. We make prediction of departure delay at predicting time, using data collected during observation window. The predicting time is set at a set duration before the gate-out time. This duration is named delay-predicting gap.} \label{fig_flight}
\end{figure}

Fig.\ref{fig_flight} illustrates the issue of departure delay by showing its effects on the optimal traffic flow. An aeroplane's arrival to departure work-flow mainly consists of five stages: airborne stage, taxi-in stage, turn around stage, taxi-out stage and airborne again. A tarmac delay occurs when an aeroplane on the ground is either awaiting takeoff (departure delay) or has just landed and passengers do not have the opportunity to get off the plane (arrival delay) \cite{Tarmac_Delays}. Compared with waiting time in the aeroplane, departure delay time, the time between the scheduled gate-out time and actual gate-out time is much more frequent. The ability to accurately predict departure delay is valuable because of the frequency at which it occurs, the complaints it generates from passengers, and the general lack of research into the problem.

Airlines usually increase the time between gate departure and gate arrival time in an effort to anticipate potential delays due to weather conditions or airport and airspace congestion. Although this operation provides passengers' additional certainty, it also increases passengers waiting time \cite{GAO19}. A more accurate departure delay prediction model can provide passengers the same level of certainty, without arbitrarily increasing waiting time.

According to the USA Department of Transportation's data (DOT), the primary cause of flight departure delay was bad weather \cite{GAO19} and traffic control. For example, the report claims that airports can be closed due to severe weather. Additionally, the number of aircraft that can be safely accommodated in a given portion of airspace further affects capacity. Overloaded airspace is likely to lead to delays on the ground or en route \cite{GAO10}.

Many previous studies discussed the correlation between weather, airport traffic and flight departure delay. Most works use the existing attributes of flight data (day of a week, season, month, etc.) and weather conditions (wind direction, wind speed, etc.) \cite{Kim2016}\cite{Sternberg2017}. Most computer scientists explore different models such as probabilistic models \cite{ahmadbeygi2010decreasing}\cite{tu2008estimating}, network representation \cite{abdelghany2004model}\cite{ahmadbeygi2008analysis}, and machine learning models \cite{rebollo2012characterization}\cite{khanmohammadi2014systems} to predict flight delay and regard it as a regression or classification problem. However, very few works explored the airport traffic situations and predicted the departure delay using aircraft trajectory data. In fact, there are only few works that works with on-ground aircraft trajeoctry data in general \cite{prabowo2019coltrane}. As the GAO report mentioned, both weather and traffic congestion at the airport results in flight delays \cite{GAO19}.

In the aerospace area, researchers propose a concept called air traffic complexity (ATC) to describe airport traffic. However, most works propose equations based on empirical observations, without offering any quantification of the concept. In this paper, we explore the characteristics of flight data, weather data, and airport traffic situational awareness data. We applied factor analysis to weather data and extract the principal components of weather features at the airport. Moreover, to our best knowledge, we are the first group to propose a new feature called Air Traffic Complexity (ATC) using airport situational awareness maps (trajectory of flights and vehicles at the airport).
% We propose new features from ATC and conduct an extensive feature importance comparison experiments to determine the their significance.
Our experiments show that proposed features from the airport situational awareness map have a significant impact on departure delay of flights, and they are even more critical than weather conditions in prediction. Moreover, using the combination of features extracted from weather conditions, historical data, our proposed features extracted from airport situational awareness map and state-of-the-art machine learning techniques, we obtained an accepted flight departure time prediction result ( $\leq$ 20 minutes error).

The contribution of this paper is summarised as follows:
\begin{itemize}
\item We propose a generic framework to predict the departure delay time using multiple data sources.
\item We propose new features extracted from the airport situational awareness map. Those features have a significant impact on departure delay prediction. We also analyse the importance of those features and compare them against standard delay-causing factors causes delays such as weather conditions.
\item We are the first group to use airport flight situational awareness maps to define the air traffic complexity (ATC).
\item We conduct extensive experiments with a large real-world dataset, including a comparison with different machine learning methods and data sources. 
\end{itemize}

The paper is organised as follows: Some methods and techniques used are shown in Section \ref{chapter:methodology}; Section \ref{chapter:ExperimentAndResults} shows the experiments and comparison studies; Section \ref{chapter:Discussion} discuss the current and future work, and Section \ref{chapter:conclusion} concludes the paper.

\section{Methodology}\label{chapter:methodology}
In this section, we will show the flight departure delay time prediction framework and discuss the details of the data pipeline.
% Fig.\ref{fig_system} depicts each component of the prediction framework.
In the first stage, we collect historical data, weather condition data, and tarmac aircraft and vehicles GPS data from different data sources. Since the collected data is noisy, incomplete and redundant, we apply the methods used in \cite{Percom2019shao} to clean and pre-process the data. 

In the feature extraction stage, we applied principal component analysis to weather data, and extracted ATC features from tarmac aircraft and vehicle trajectory data. We also utilise the historical scheduling table data. In the modelling stage, we combined multiple datasets and use various data combinations to train a regressor model that can be used for predicting departure delay time. To show the impact of integrating our novel spatial features, in the form of ATC, we choose four popular regressors from different families (linear regression, SVR, ANN, and regression trees) to show the robustness of our proposed approach to different regressors.

% \begin{figure*}
% \includegraphics[width=0.9\textwidth]{graph/Sys_Arch3.pdf}
% \caption{The demonstration of the overall system architecture.} \label{fig_system}
% \end{figure*}

% \subsection{Data Pre-processing}\label{sec_Data_cleaning}
% \begin{figure*}[htbp]
% \centering
% \subfigure[Three types of Tarmac area]{
% \includegraphics[width=0.45\linewidth]{graph/LAX-withArrow.png}
% }
% \subfigure[Heatmap of the aircraft position]{
% \includegraphics[width=0.45\linewidth]{graph/takeoff-Density.png}
% }
% \caption{The tarmac area is classified into 3 types: apron/runway/parking area.}
% \label{fig_tarmacmap}
% \end{figure*}

The raw trajectories GPS data is noisy and contains some irrelevant data such as the GPS tracks for non-aircraft vehicles. As such, these noises are removed in the pre-processing step. Examples of noise include the absent altitude value and the GPS points that do not fall within the possible continuous trajectories of a particular aircraft. Since the same flight on different dates will share the same call sign, we create a data pipeline to filter and categorise the aircraft trajectories based on date and the call sign of the flight.

The data pipeline also calculates the speed of the aircraft, using time and distance deltas, and labels the trajectories based on the tarmac areas they cross. We classify the tarmac area into three different types: parking area, apron area, and runway area. %An example of these areas are shown in Fig.\ref{fig_tarmacmap}.

\subsection{Feature Extraction}
Considering the delay propagation effect, we match the departure flights with their previous records using historical data, which contain the schedule/actual arrival time for each aircraft's previous flight. For the weather features, we use the most recent weather data within the observation window.

In order to deal with the high-dimensional weather data, we performed a principal component analysis (PCA) on it. The PCA is a statistical procedure which is applied to a set of high-dimensional, correlated attributes and transformed them into another set of variables that are linearly uncorrelated \cite{li2014principal}. This process reduces the dimension of the feature set while retaining the knowledge in the original set. We extracted 18 principal components from weather data using PCA.
 
For ATC data, we extract features from aircraft and vehicle trajectories on the tarmac before the prediction time. As shown in Fig.\ref{fig_flight}, there is a duration period called observation window. During this window, we extracted features from ATC data and their descriptions. The features that we extract cover the traffic density in all three tarmac area areas, as previously defined,  and the number of potential landing/take-off aircraft is also accounted for by the ATC.

Since departure delay prediction is a clear regression task, we use the square Root Mean Squared Error (RMSE) as the loss function.
% Additionally, we use $R^2$ score as a secondary metric, which will support the result of RMSE.  

\section{Experiment and Results} \label{chapter:ExperimentAndResults}
\subsection{Datasets}
\subsubsection{Schedule Table Data}

The flight info data is an open dataset collected by the Bureau of Transportation Statistics of United  State  Department of Transportation \cite{historicaldata}. It consists of the flight records for LAX from the 1\textsuperscript{st} of July, to the 31\textsuperscript{st} of August, 2016. It covers both the arrival and departure flight records and contains many useful attributes. Most relevant is Departure Delay, which becomes the variable we are attempting to predict in this paper. Additionally, the Scheduled Arrival/Departure time and the Cause of Delay (with is categorised in 5 different types) which are used as features after pre-processing, and the Call-sign/Tail-number which helps us to link the aircraft and flights.

\subsubsection{Weather Data}

The weather data in this experiment is collected from website \cite{weathersource}. This daily weather observation in the Los Angeles Airport (LAX) is captured every hour and it covers most aspects except the basic temperature and humidity values, such as the wind direction and wind speed and ambient air pressure values.

% \begin{table*}[htbp]
% \caption{The attributes in the weather dataset}\label{tab:weather_attributes}
% \centering
% \begin{tabular}{p{0.2\textwidth}|p{0.72\textwidth}}
% \hline 
%  \textbf{  Attribute Name} & \textbf{  Description} \\
% \hline 
% \hline
%  Time \ \ \  & The timestamp of when this data been captured. \\
% \hline 
%  Temperature & The actual temperature at LAX (Fahrenheit). \\
% \hline 
%  Dew Point & The temperature of dew point. \\
% \hline 
%  Humidity & Humidity at that time (\%). \\
% \hline 
%  Wind Direction & This is a categorical value about the direction of the wind (e.g. N/S/W/E/NW, etc.). \\
% \hline 
%  Wind Speed & The speed of the wind in that certain direction (mph). \\
% \hline 
%  Wind Gust & The potential increase in the speed of the wind \ (usually under 20s).  \\
% \hline 
%  Pressure & The ambient air pressure at LAX. \\
% \hline 
%  Condition & The overall weather condition at LAX which is also a categorical value (e.g. Fair/Mostly Cloudy) \\
%  \hline
% \end{tabular}
% \end{table*}

We use one-hot encoding to transform categorical features, such as Condition and Wind Direction, into numerical values, which is necessary when using decision trees.

\subsubsection{Airport GPS Trajectories Data}

The GPS trajectories are a private dataset collected by the United States' Federal Aviation Administration's (FAA's) System Wide Information Management (SWIM) program. Access to this dataset is acquired through our industry partner. Spatially, it is contained to the tarmac of LAX. Temporally, the data spans 7 weeks in July and August in 2016. It contains GPS trajectories of both aircraft and ground vehicles. There are around 11 million GPS points, grouped into 43 503 trajectories, belonging 6 518 vehicles.

\subsection{Experimental Results}

% \begin{table}
% \caption{The performance improvement .}\label{tab2}
% \centering
% \begin{tabular}{p{0.14\textwidth}|p{0.30\textwidth}|p{0.25\textwidth}|p{0.26\textwidth}}
% \hline 
%  \textbf{Model} & \textbf{Historical + Weather} & \textbf{Historical + ATC} & \textbf{Fused Data} \\
% \hline 
%  LR & 0.13\% & 0.02\% & 0.11\% \\
% \hline 
%  SVR & 0.67\% & 0.96\% & 1.12\% \\
% \hline 
%  MLP & -2.51\% & 0.01\% & 0.91\% \\
% \hline 
%  LGBM & -0.57\% & 0.76\% & 2.04\% \\
%  \hline
% \end{tabular}
% \end{table}

% In this section, we conducted three sets of experiments. In this first experiment, we evaluated the prediction results using different combinations of data sources and machine learning models. In the second experiments, we tested the temporal sensitivity of the learning model and validated the robustness of model in flight delay time prediction. In the last set of experiment, we compared the importance of different features that we extracted and created from historical data, ATC dataset, and weather dataset. We conducted these three sets of experiments to evaluate the performance of our proposed flight delay prediction framework and importance of our proposed ATC features.

We evaluated the prediction results using different combinations of data sources and machine learning models. We apply three conventional machine learning regressor Linear Regression (LR), Support Vector Regressor (SVR), Multilayer Perceptron (MLP), and one state-of-the-art boosting tree based regressor: LightGBM to predict the flight delay time using different combination of data sources (Historical data, Airport weather conditions data and ATC data). In this experiment, we use the classic evaluation metric: Root Mean Square Error (RMSE) to measure the performance of flight delay time prediction using different models and different combinations of data sources. Lower RMSE shows the less error between prediction value and true value. We compared the importance of different features that we extracted and created from historical data, ATC dataset, and weather dataset to evaluate the performance of our proposed flight delay prediction framework and importance of our proposed ATC features. 

\begin{figure}[htbp]
\centering
\subfigure[The result for RMSE]{
\includegraphics[width=0.46\textwidth]{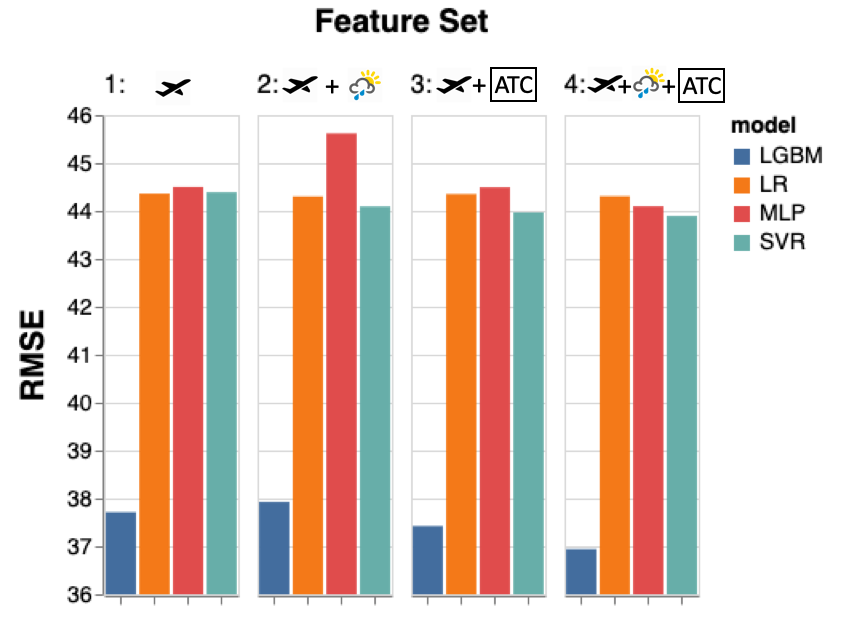}
}
% \subfigure[The result for $R^2$]{
% \includegraphics[width=0.46\textwidth]{graph/r2.png}
% }
\caption{The RMSE result for different models and different combination of different data sources.}
\label{fig_result_barchart}
\end{figure}

As shown in the Fig.\ref{fig_result_barchart}, LightGBM has the best overall performance. And we can found out that there is a clear performance boost after adding the ATC features we created. The best result comes from the dataset that combines flight arrival data, ATC features, and weather data, however, adding weather data alone with the arrival dataset did not result in a clear improvement and sometime makes the result even worse. These results indicate that the ATC features contribute more than the weather features when predicting flight departure delay.

In summary, prediction accuracy is stable with even less training data. Furthermore, even the prediction time is four hours ahead the flight delay event, the accuracy is around 18 minutes.

% In the third experiment, we compare the importance of features we used in ATC dataset, weather condition dataset and historical dataset. We use the parameter $feature\_ importance$ recorded by the LightGBM regressor, which captures the numbers of times a certain feature is used to `split' a tree \cite{featureimportance}. The higher times this value is, the more information this feature provides which helps the model to differentiate the situation. 

% \begin{figure*}
% \includegraphics[width=\textwidth]{graph/foocopy.pdf}
% \caption{The feature importance for Weather and ATC features. All ATC features in this graph has a red asterisk before them.} \label{fig_feature_importance}
% \end{figure*}

% Fig.\ref{fig_feature_importance} shows the feature importance after the training process. Firstly, the historical data is still the most important feature (Schedule Departure Time) in predicting departure delay. Secondly, we can find that most of our selected ATC features have a significant higher importance than the weather features, which also supports the result in the first experiment that the ATC features does contribute more in this task.

\section{Discussion and Future Work}\label{chapter:Discussion}
Prior work has documented the effectiveness of weather conditions and air traffic in improving the accuracy of flight delay time prediction. However, these studies have either been arrival delay prediction or have not focused on analysing trajectory data of flights on the tarmac. In this study, we pay attention to trajectories of aircraft and vehicles at the airport and use those features to convert an abstract concept air traffic complexity (ATC) to a concrete concept which can be formalised by aircraft and vehicle trajectory data.  We found that in our extensive comparison experiments, features extracted from ATC are significantly crucial to departure delay which is more influential in air passengers. These findings suggest that air traffic controllers and airlines should spend more efforts on airport vehicles and aircraft arrangements. Compared with weather conditions, ATC factors are more controllable and easy to change.
Additionally, departure delay prediction is more critical than arrival delay to passengers, and passengers usually conduct more complaints about tarmac delay because human beings dislike the uncertainty. Therefore, we develop a prediction model on departure delay prediction time only. Most notably, this is the first study to our knowledge to investigate the importance of airport situation awareness map in departure delay prediction. Our results provide compelling evidence.

However, although our hypotheses were supported statistically, there remain many limitations and potential work in the future.

Firstly, we only focus on the departure delay time prediction in this paper. Future work should, therefore, include follow-up work designed to evaluate whether the proposed features also play an essential role in other types of flight delay prediction. Secondly, except for the features we extracted, more potentially useful features can be explored from the airport situational awareness map. Thirdly, for each feature, we can still apply the sensitivity analytic and provide useful suggestions to the traffic controller to manage the airport. Finally, an experiment includes multiple airport data should be used to generalise our results and validate our assumption in the future.

\section{Conclusion}\label{chapter:conclusion}
In this paper, we have developed a data-driven framework to predict the departure flight delay time and explored the importance of different features extracted from the airport situational awareness map. We have demonstrated the airport situational awareness map plays a more important role in departure delay time prediction than weather conditions. We have demonstrated that LightGBM regressor outperforms other conventional regressors with extensive experiments on a large real-world dataset. The code and sample data will be published online in the future.

\begin{acks}
This research is funded by Northrop Grumman Corporations USA for "Spatio-temporal Analytics for Situation Awareness in Airport Operations" project, RMIT University.
\end{acks}

\bibliographystyle{ACM-Reference-Format}
\bibliography{mybibliography}

%\appendix

\end{document}